\documentclass[aps,prb,twocolumn,showpacs,showkeys,superscriptaddress]{revtex4-1}

\usepackage[pdftex]{graphicx}
\usepackage{bm}

\begin{document}
\title{Effect of current injection into thin-film Josephson junctions}


\author{V. G. Kogan} 
\email[]{kogan@ameslab.gov} 
\affiliation{  Ames Laboratory, US Department of Energy, Ames, Iowa 50011, USA }
 \author{R. G. Mints}
\email[]{mints@post.tau.ac.il}
\affiliation{The Raymond and Beverly Sackler School of Physics and
Astronomy, Tel Aviv University, Tel Aviv 69978, Israel}

 \begin{abstract}
New thin-film  Josephson junctions have recently been  tested 
in which the current injected into one of the junction banks  governs Josephson phenomena. One thus can continuously manage the phase distribution at the junction by changing the injected current. A method of calculating the distribution of injected 
currents is proposed for a half-infinite thin-film strip with source-sink points  at arbitrary positions at the film edges. 
The strip width $W$ is assumed small relative to $\Lambda=2\lambda^2/d$, $\lambda$ is the bulk London penetration depth of the film material, $d$ is the film thickness.  
\end{abstract}



\maketitle
%

  \section{Introduction } 
  In  recent  years, the physics of Josephson phenomena enjoyed  a number of important developments. Introduction of $\pi$ and  0-$\pi$ junctions, \cite{Lev}   various ways to have a different from $\pi$ phase shift, \cite{Kirtley,Mints-phi}  effect of vortices in the junction vicinity, \cite{Krasnov,KM1,KM2}  to name a few. Striking improvements in managing Josephson phenomena  came after   injection of  currents into one of the thin-film banks was introduced that allowed for continuous control of the phase difference on the junction \cite{Ust,Gold1,Gold3} 
  and, in particular, to imitate the  0-$\pi$ behavior.  
 This  development   necessitates  evaluation of the injected supercurrent distribution in one of the junction banks since this determines the distribution of the superconducting phase. 
 
  \begin{figure}[h]
\begin{center}
 \includegraphics[width=7.cm] {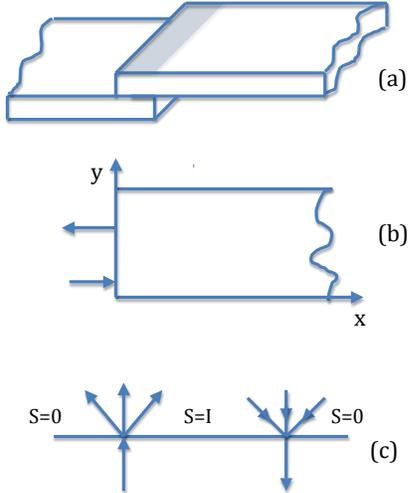}
\vskip -2 cm
\caption{(a) A sketch of two semi-infinite thin-film strips forming the Josephson junction in the overlapping shaded region. (b) The top film; arrows show  positions of injection point contacts. (c) The stream function $S$ at the film edge. 
}
\label{f1}
\end{center}
\end{figure}

 In one of  common realizations, the junction is formed by two ``half-infinite" thin-film strips with overlapped edges. Extra current injectors are attached to the edges of one of the films, shown schematically in Fig.\,\ref{f1}. The injected  current  affects the phase distribution in the thin-film bank where it flows and thus the phase difference on the junction.
  We show in this communication that for sufficiently thin films with the size $W$ smaller than the Pearl length $\Lambda=2\lambda^2/d$ the problem of the injected currents  can be solved under   very general assumptions, so that the design of junctions with needed properties becomes possible. 

 \section{Stream function}
  Consider a half-infinite thin-film strip of a width  $W \ll \Lambda=2\lambda^2/d$  where $\lambda$ is the London penetration depth of the film material and $d$ is the film thickness. Choose $x$ along the strip, $0<x<\infty$, and $y$ across so that $0<y<W$, Fig.\,\ref{f1}b. Let the injection   points be at $ (x_1, y_1)$ and $ (x_2, y_2)$ at the film edge.  

The London equation integrated  over the film thickness  reads:
\begin{equation}
h_z + {2\pi\Lambda\over c}\,{\rm curl}_z\,{\bm g}= 0\,.
\label{London}
\end{equation}
Here,   ${\bm g}$ is  the sheet current density and $h$ is the self-field of the current $\bm g$. The Biot-Savart integral for $h_z$ in terms of $\bm g$ shows that $h_z$  is of the order $g/c$, whereas the second term on the left-hand side of
Eq.\,(\ref{London}) is of the order $g\Lambda/cW\gg g/c$. Hence, in narrow strips with $W\ll \Lambda$, the self-field can be disregarded.  Introducing the scalar stream function $S$ via $\bm g = {\rm curl}[ S(x,y)\hat z]$, we obtain instead of Eq.\,(\ref{London}):
\begin{equation}
\nabla^2 S = 0 \,.
\label{Laplace}
\end{equation}
  Physically, this simplification comes about since in narrow films the major contribution to the system energy is the kinetic energy of supercurrents, while  their magnetic energy can be disregarded.  

 The boundary condition of zero  current component normal to edges, e.g., $g_y=-\partial_xS=0$ at the edge $y=0$,  translates to $S=$ constant along  the edges.  This constant, however, is not necessarily the same everywhere, in particular, it should experience a finite jump at injection points. Consider  a point contact at the edge  as illustrated at Fig.\,\ref{f1}, take its position as the origin of polar coordinates,  and integrate the $r$ component of the current $\bm g$ along the small half-circle centered at the injection point. The total injected current is:
  \begin{eqnarray}
I=\int_0^\pi g_r \,r\,d\phi = \int_0^\pi \frac{\partial S}{\partial\phi} \,d\phi = S(\pi)-S(0)\,.
 \label{I}
\end{eqnarray}
Hence, on two sides of the injection point the stream function experiences a jump equal to the total injected current. Clearly, at the current sink  $S$-jump has the opposite sign. Thus, we can choose $S=0$ everywhere at the edges of the half-infinite strip, except the segment between the injection and sink points, where $S=I$. 

To solve the Laplace equation (\ref{Laplace}) we first employ the conformal mapping of the half-strip to a half-plane: \cite{Morse,KM2}   
 \begin{eqnarray}
u+i v=-i\cosh \pi(x+i y)\,. 
 \label{conf}
\end{eqnarray}
It is seen that   the   half-plane $u>0$ is transformed to the half-strip of a width 1 (hereafter we use $W$ as a unit length).  Explicitly, this transformation  reads:
 \begin{eqnarray}
u =\sinh \pi x\,\sin \pi y  \,, \qquad  v=-\cosh \pi x\,\cos \pi y\,.  
 \label{uv}
\end{eqnarray}
 Hence, we have to solve the Laplace equation on a half-plane $u>0$ subject to boundary conditions $S=I$ at the edge $u=0$ in the interval $v_1<v<v_2$ with 
  \begin{eqnarray}
v_1=-\cosh \pi x_1\cos\pi y_1\,,\quad v_2= -\cosh \pi x_2\cos \pi y_2 \,,\qquad
 \label{interval}
\end{eqnarray}
 and $S=0$ otherwise.
  
 To proceed, we first write the ``step-function" $S(0,v)$ at the edge $u=0$ as a Fourier integral:
  \begin{eqnarray}
S(0,v) &=&\int_{-\infty}^\infty \frac{dk}{2\pi} S(0,k)e^{ikv}\,,\label{Fourier}\\
 S(0,k) &=&I\int_{v_1}^{v_2}dv\,e^{-ikv}=\frac{I}{ik}\left(e^{-ikv_1}-e^{-ikv_2}\right).
 \label{Fourier-comp}
\end{eqnarray}
Since $S(0,v) $ is a linear superposition of plane waves $e^{ikv}$, we first consider the solution of the Laplace equation $(\partial^2_u  +   \partial^2_v) s(u,v) =0$ subject to the  boundary condition $s(0,v)=e^{ikv}$. Separating variables   we  obtain  $s(u,v)  = e^{ikv}\,e^{-|k|u}$.
 Hence, the solution for the actual boundary condition is:
  \begin{eqnarray}
S(u,v) =\int_{-\infty}^\infty \frac{dk}{2\pi} S(0,k)e^{ikv-|k|u}\,.
 \label{solution}
\end{eqnarray}
Substituting here $S(0,k)$ of Eq.\,(\ref{Fourier-comp}) one obtains:
  \begin{eqnarray}
S(u,v) =\frac{I}{\pi}\left(\tan^{-1}\frac{v-v_1}{u}-\tan^{-1}\frac{v-v_2}{u}\right)\,.
 \label{solution1}
\end{eqnarray}
It is seen that at $u\to 0$,  $S=I$ if $v_1<v<v_2$ and $S=0$ othewise, as it should be.
 Thus, we have the steam function at the half-plane $u>0$ for arbitrary positions $v_1$ and $v_2$ of the current contacts at the edge $u=0$. 
  We now can go back to the $(x,y)$ plane and specify the injection positions. 
 
 \subsection{Injectors at the edge $\bm {x=0}$}
\begin{figure}[h]
\begin{center}
 \includegraphics[width=8.cm] {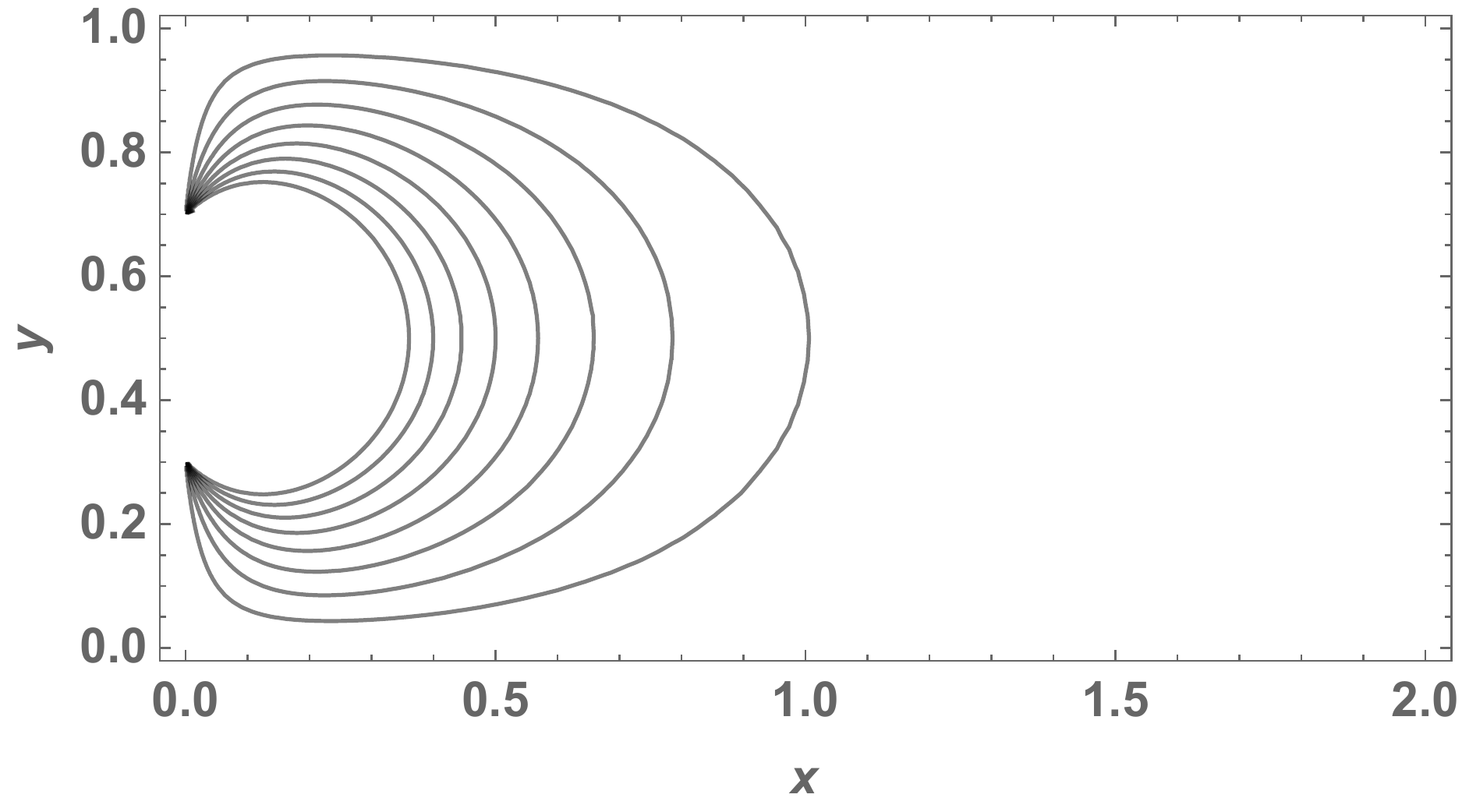}
\caption{ The current distribution for the injection points   $(0,0.3)$ and   $(0,0.7)$.  
}
\label{f2}
\end{center}
\end{figure}

 Let the injector and the sink be at $(0,y_1)$ and $(0,y_2)$.  We obtain:
   \begin{eqnarray}
\frac{S(x,y)}{I/\pi} &=& \tan^{-1}\frac{\cos \pi y_1-\cosh \pi x\cos \pi y}{\sinh \pi x\,\sin \pi y}\nonumber\\
&-&\tan^{-1}\frac{\cos \pi y_2-\cosh \pi x\cos \pi y}{\sinh \pi x\,\sin \pi y} \,.
 \label{solution2}
\end{eqnarray}
The lines of the current $\bm g$ are given by $\bm g\times d\bm r=0$ or by $\partial_xS\,dx+\partial_yS\,dy=dS=0$, in other words, by contours of $S=$ const.
An example is shown in Fig.\,\ref{f2}.
\vskip -1 cm

 \subsection{The injector  at  $\bm {y=0}$ and the sink at $\bm {x=0}$}
\begin{figure}[h]
\begin{center}
 \includegraphics[width=8.cm] {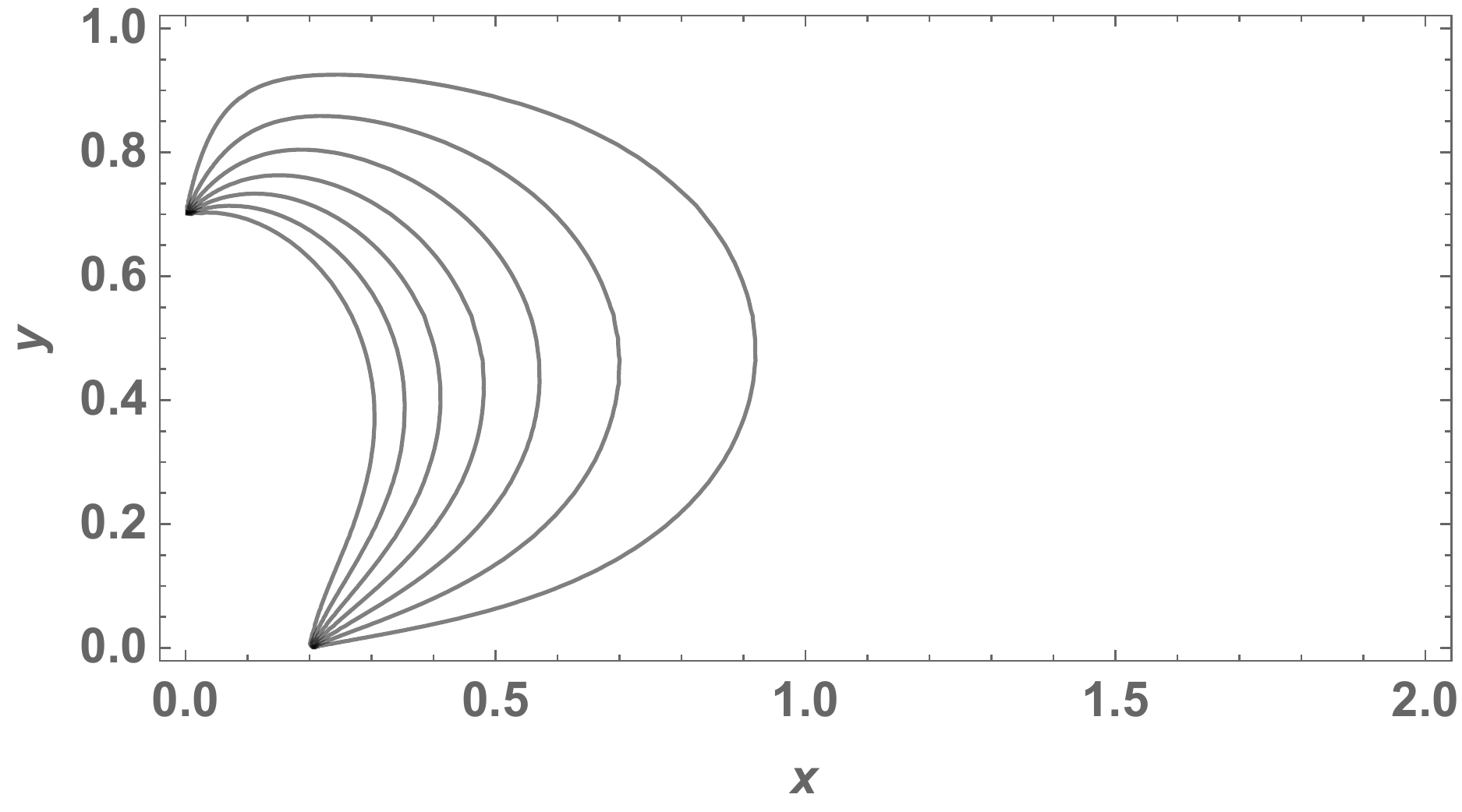}
\caption{ The current distribution for the asymmetric injection points   $(0.2,0)$ and   $(0,0.7 )$.  
}
\label{f3}
\end{center}
\end{figure}

 This situation corresponds to positions $(x_1,0)$ and $(0,y_1)$:
   \begin{eqnarray}
\frac{S(x,y)}{I/\pi} &=& \tan^{-1}\frac{\cosh \pi x_1-\cosh \pi x\cos \pi y}{\sinh \pi x\,\sin \pi y}\nonumber\\
&-&\tan^{-1}\frac{\cosh \pi y_2-\cosh \pi x\cos \pi y}{\sinh \pi x\,\sin \pi y} \,.
 \label{solution4}
\end{eqnarray}
An example is shown in Fig.\,\ref{f3}.
 
 It is worth noting that the same method can be employed for  currents injected to   thin-film samples of any polygonal shape. According to the Schwartz-Christoffel  theorem any polygon can be mapped   onto a half-plane. The general solution (\ref{solution1}) on the $(u,v)$ plane will  hold. Therefore, knowing the function which realizes the needed transformation, one obtains $S(x,y)$. The only physical precondition for this method to work is the requirement  of a small sample size on the scale of Pearl length $2\lambda^2/d$, that allows one to reduce the problem to the Laplace equation for the stream function $S$. The method can be applied for more than two injection points or to extended injections, which require though different boundary conditions imposed on $S$.
 
 \section{Phase} 
We now note that the sheet current is expressed either in terms of the gauge invariant phase $\varphi$ or via the stream function $S$:  
$${\bm g}=-\frac{c\phi_0}{4\pi^2\Lambda}\,\nabla \varphi 
={\rm curl}\,S {\bm z}$$  
This  relation written in components shows that 
$S({\bm r})$ and $(c\phi_0/4\pi^2\Lambda)\varphi({\bm r})$
are the real and imaginary parts of an analytic function.   

It is easy to construct the phase on the $(u,v)$ plane since   
$$-i\ln (u+i v)=\tan^{-1}(v/u) - i\ln\sqrt{u^2+v^2}.$$ 
 Hence, the phase corresponding to the  stream function of Eq.\,(\ref{solution1}) obeys:
   \begin{eqnarray}
\frac{c\phi_0}{4\pi^2\Lambda}\,\varphi =-\frac{I}{\pi}\ln\frac{ \sqrt{u^2+(v-v_1)^2}} {\sqrt{u^2+(v-v_2)^2}}\,,
 \label{phase0}
\end{eqnarray}
or
   \begin{eqnarray}
\varphi =-\frac{I}{I_0}\ln \frac{u^2+(v-v_1)^2}  { u^2+(v-v_2)^2}\,,\quad I_0=\frac{c\phi_0}{4\pi \Lambda}\,.
 \label{phase}
\end{eqnarray}
The characteristic current $I_0$ depends on the Pearl length. 
Thus, the phase is proportional to the reduced injected current  $j=I/I_0$ and a factor $\varphi/j$  depending on the film geometry and injection positions. Substituting here $u(x,y)$ and $v(x,y)$ one obtains the phase as a function of $(x,y)$. 

\begin{figure}[h]
\begin{center}
\includegraphics[width=8cm] {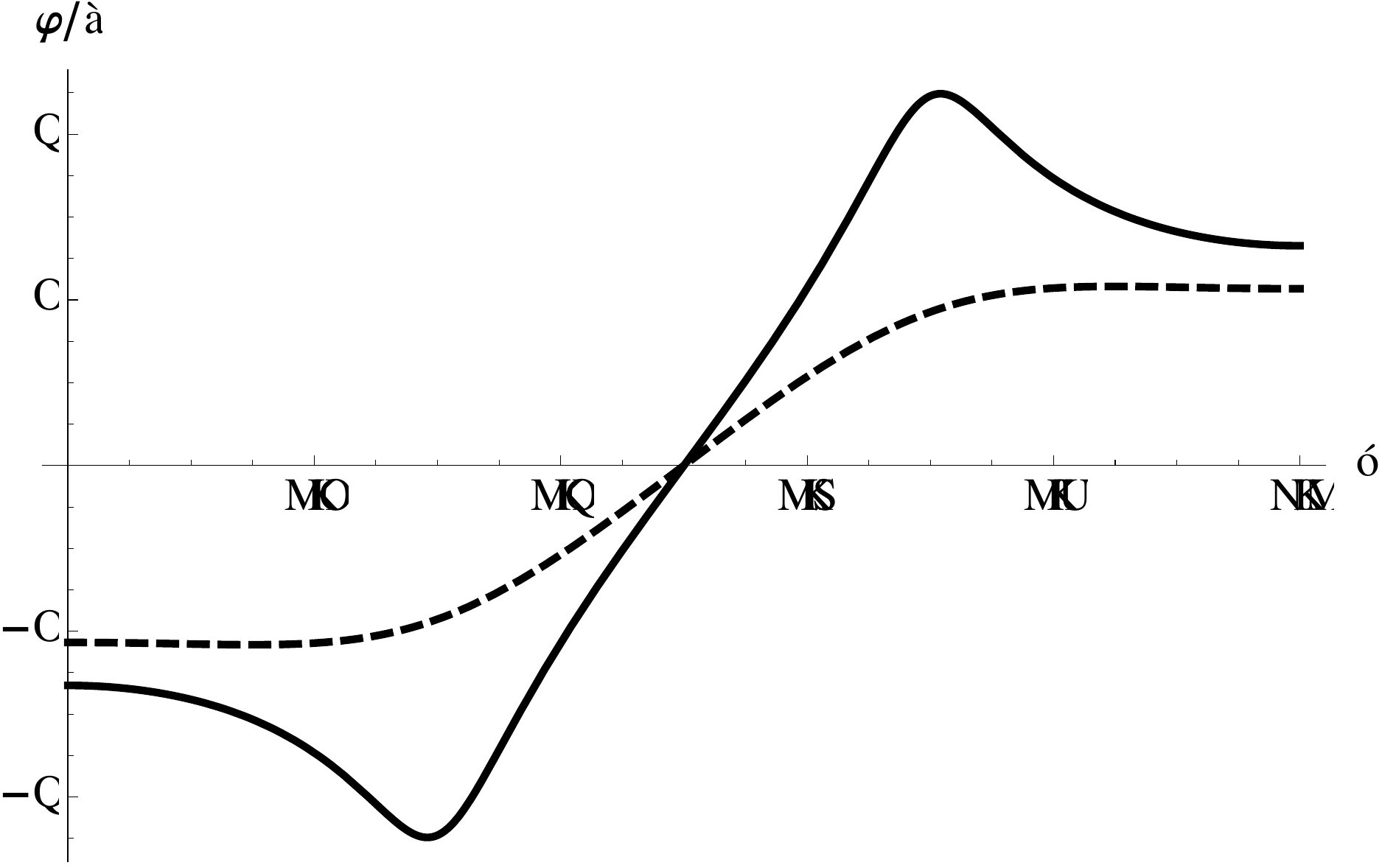} 
\caption{The phase $\varphi(x_0,y)$ for $j=1$ at fixed distances $x_0$ from the edge $ x=0 $ as a function of the transverse coordinate $y$ for symmetric contacts at     $y_1=0.3$ and $y_2=0.7$. The solid line is for $x_0=0.05$, the dashed line is for $x_0=0.2$. }
\label{f3a}
\end{center}
\end{figure}

Examples of the phase near the edge $x=0$ are shown in Fig.\,\ref{f3a}. 
Clearly, one can make the phase ``jump" steeper by putting injection contacts closer. It is worth noting that  the phase change due to injected currents can be used, e.g., to imitate properties of 0-$\pi$ junctions, or in fact to have any phase shift by choosing properly the injected current. 

 \section{Josephson critical current}
  The total Josephson current through a rectangular patch (the shaded region in Fig.\,\ref{f1}a) with the size $\Delta x$ along the $x$ axis is
    \begin{eqnarray}
\frac{J(I/I_0)}{J_{c0}}=\int_0^{\Delta x}dx \int_0^1 dy\sin[\varphi(x,y,I/I_0)+\varphi_0]\,,\qquad
 \label{int}
\end{eqnarray}
where $J_{c0}$ is the critical Josephson current density (which in the following is set equal to 1) and  $\varphi_0$ is an overall phase imposed by the transport current through the junction. 
To find the critical current, we maximize this relative to $\varphi_0$ to obtain $J_c =\sqrt{A^2+B^2}$ where 
    \begin{eqnarray}
A=\int_0^{\Delta x} dx \int_0^1 dy\sin[\varphi(x,y,;j) ]\,,\nonumber\\
 B=\int_0^{\Delta x}dx \int_0^1 dy\cos[\varphi(x,y;j) ]\,.
 \label{AB}
\end{eqnarray}
  
 \subsection{Symmetric injection}
\begin{figure}[h]
\begin{center}
\includegraphics[width=8.cm] {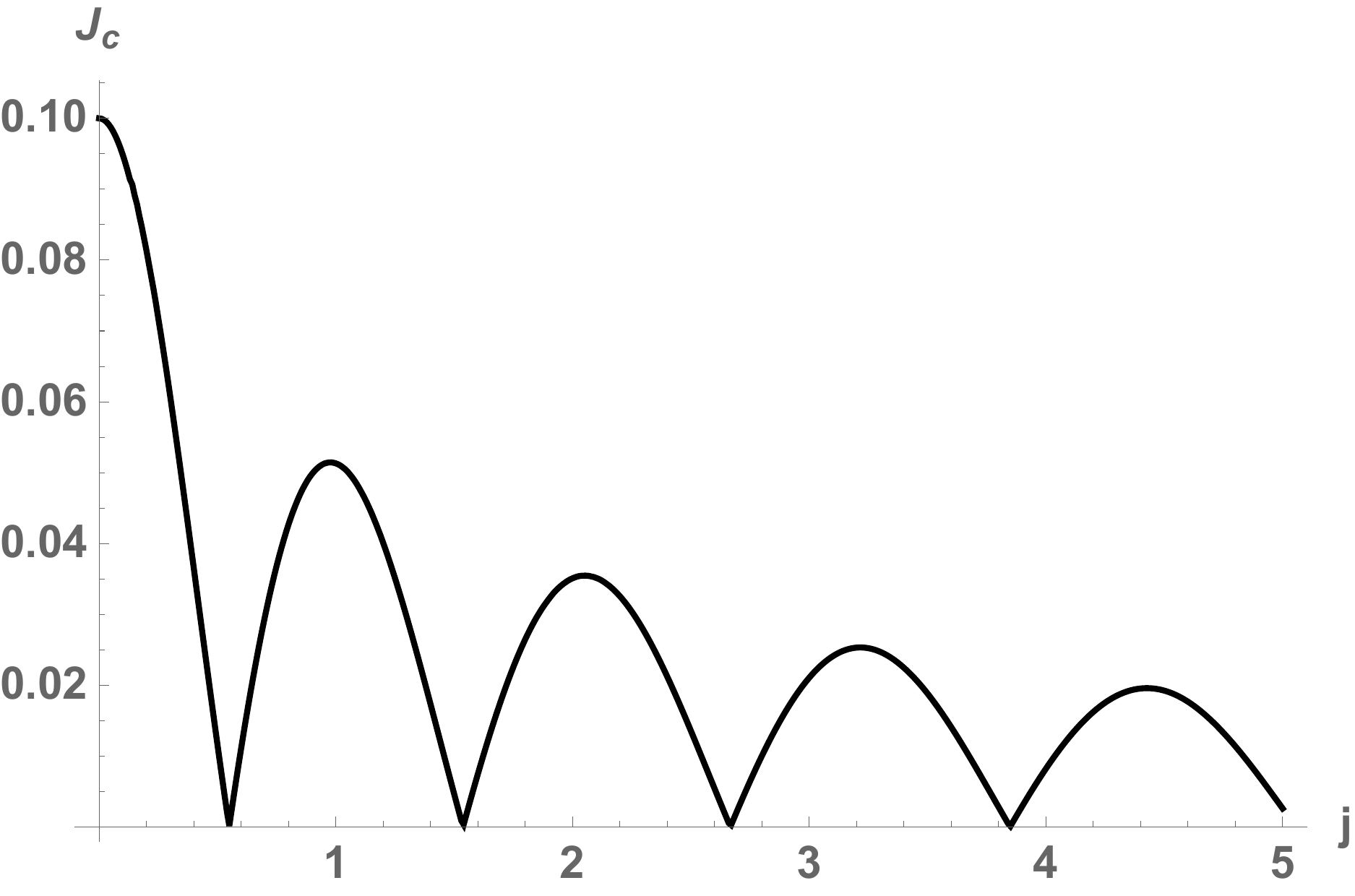} 
\caption{$J_c(j)$ for the junction width $\Delta x=0.1$  and symmetric contacts at  the edge $x=0$.  $y_1=0.3$ and $y_2=0.7$ so that the distance between contacts is $\Delta y=0.4$.  }
\label{f5}
\end{center}
\end{figure}

Consider  $J_c(j)$ for the rectangular junction of the width $\Delta x=0.1\,W$,  similar  to the  experimental set up. \cite{privat}  The injection contacts are at the edge $x=0$ and $y_1=0.3$ and $y_2=0.7$, i.e. they are symmetric relative to the strip middle $y=1/2$. The current distribution for this case is given in Fig.\,\ref{f2}. $J_c(j)$ evaluated with the help of Eq.\,(\ref{AB}) is shown in Fig.\,\ref{f5}.
We note  that $J_c(0)$ is  proportional to the junction width $\Delta x$,   since the Josephson critical current density is constant in the absence of injected currents.

\begin{figure}[h]
\begin{center}
\includegraphics[width=8.cm] {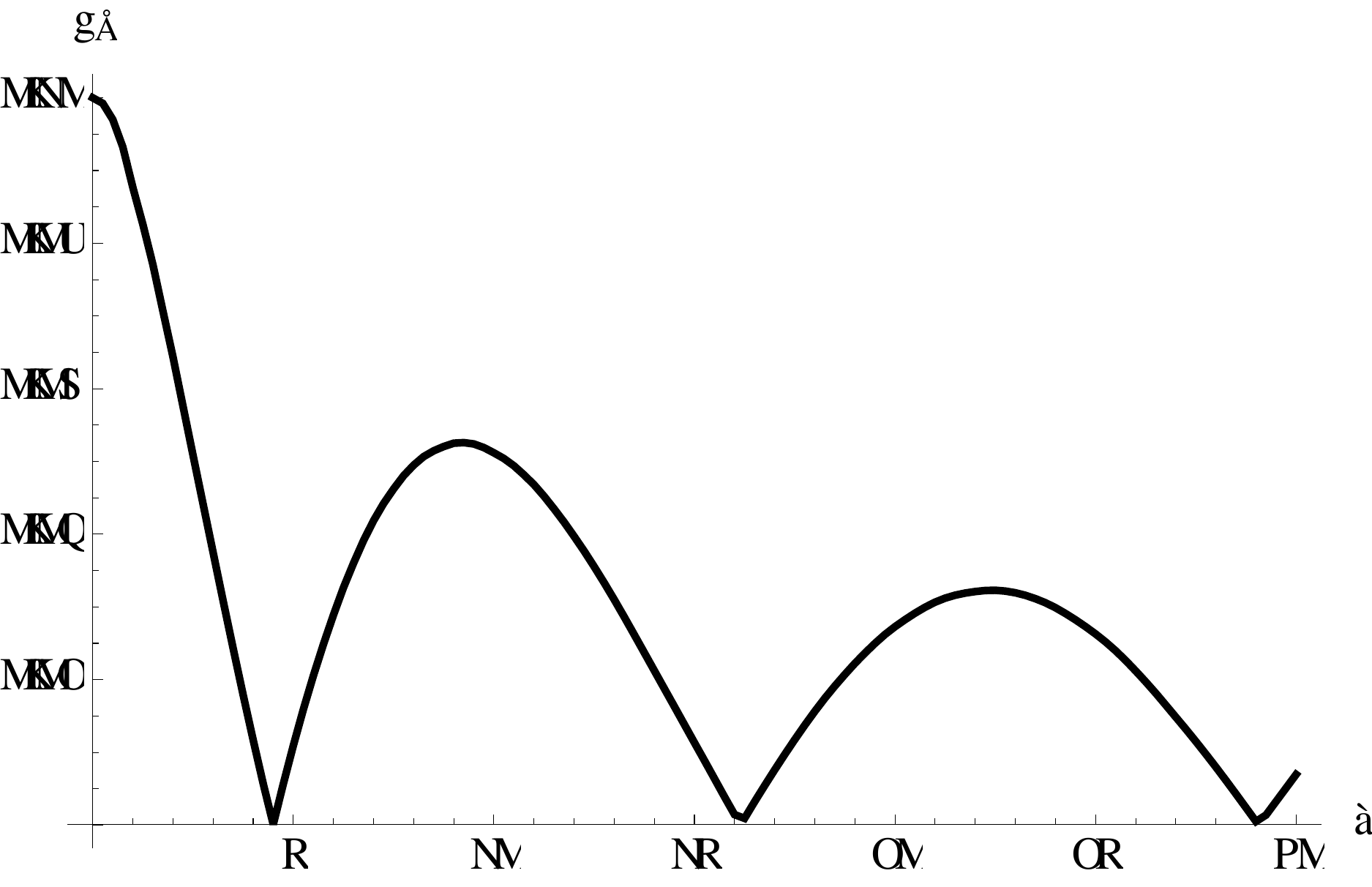} 
\caption{$J_c(I/I_0)$ for the junction width $\Delta x=0.1$  and symmetric contacts at  the edge $x=0$ and $y_1=0.48$ and $y_2=0.52$ so that the distance between contacts is $\Delta y=0.04$.   }
\label{f6}
\end{center}
\end{figure}

Fig.\,\ref{f6} shows $J_c(j)$ in the same junction with contacts at $y_1=0.48$ and $y_2=0.52$ so that they are separated by $\Delta y=0.04$, ten times closer than in the previous example.
Comparing these plots we see that the first zero of $J_c$ at $j\approx 0.5$ in the first graph whereas it is at $j \approx  5$ in the second. We then conclude that zeros roughly scale as the inverse of the contact separation, $1/\Delta y$.  
We also observe that maxima of $J_c$ seem to be independent of contacts separation. It is shown below that these properties of $J_c(j)$ can be traced back to  general expressions (\ref{int}) and (\ref{AB}) for narrow junctions, $\Delta x\ll 1$,  and small contacts separations $\Delta y\ll 1$. 

To this end, we note that being  a solution of the Laplace equation for a half-infinite strip, the phase  $\varphi(x,y)$    changes considerably  on distances of the order $W=1$ from the edge $x=0$ where the contacts are placed. Hence, for narrow junctions with $\Delta x\ll 1$, one can set in the first approximation:   
   \begin{eqnarray}
\varphi(x,y) \approx \varphi(0,y)=-j\ln \frac{ (v-v_1)^2}  {  (v-v_2)^2}\nonumber\\
= -j\ln \frac{ (\cos \pi y-\cos \pi y_1)^2}  {  (\cos \pi y-\cos \pi y_2)^2} \,.
 \label{ph_appr}
\end{eqnarray}
Here, $y_{2,1}=(1\pm\Delta y)/2$, and for $\Delta y\ll 1$ one can expand the last expression in powers of $\Delta y $:
   \begin{eqnarray}
 \varphi(0,y)=  \frac{ 2\pi j\Delta y}  { \cos\pi y} +{\cal O}(\Delta y)^3  \,.
 \label{ph_1}
\end{eqnarray}
Since $ \cos\pi y$ is odd relative to the strip middle, and so is $\sin \varphi(0,y)$, we have  $A=0$, and $J_c=|B|$, see Eq.\,(\ref{AB}):
    \begin{eqnarray}
 J_c&=&\Big| \int_0^{\Delta x} \int_0^1 dx\,dy\cos[\varphi(x,y) ]\Big| \nonumber\\
 &=& \Delta x  \Big|  \int_0^1  dy\cos\frac{2\pi j\Delta y}{\cos\pi y}\Big|\,. 
  \label{Jc-sym}
\end{eqnarray}
Clearly, when no current is injected, $J_c(0)=\Delta x$ as is seen in Figs.\,\ref{f5} and \ref{f6}. 
The integral over $y$   can be done:
    \begin{eqnarray}
 \frac{J_c}{\Delta x}= \left |1-\eta\,\,_pF_q\left[\left\{\frac{1}{2}\right\}, \left\{1,\frac{3}{2}\right\},-\frac{\eta^2}{4}     \right]\right | \,,
    \label{pFq}
\end{eqnarray}
 where $_pF_q$ is the generalized hypergeometric function and $\eta=2\pi j \Delta y$. This function   is expressed in terms of Bessel and Struve functions so that it  oscillates  when $\eta$ changes:
    \begin{eqnarray}
   J_0(\eta) -\frac{\pi  }{2}\,\left[ J_0(\eta)\bm{H}_1(\eta)-  J_1(\eta)  \bm{H}_0(\eta)\right].
    \label{pFq1}
\end{eqnarray}
 
It is worth noting that  $J_c$ of Eqs.\,(\ref{Jc-sym}) and (\ref{pFq})   depends on the injected current $j$ and the contacts separation $\Delta y$ only via the product  $\eta=2\pi j \Delta y$. 
In other words, the curves $J_c(j,\Delta y)$ for different contact separations $ \Delta y $ can be rescaled to a universal curve $J_c( \eta)$. In particular, the zeros of $J_c(j)$ and the positions of its maxima should scale as $1/\Delta y$. Unlike their positions, the  absolute value  of its $n$-th maximum is  independent  of the separation $\Delta y$.
 
The two first roots of $J_c(\eta)$ found numerically are 1.11 and 4.06. For $ \Delta y=0.4$, we then obtain  the two first zeros of $J_c(j)$: $j_1=\eta_1/2\pi\Delta y\approx 0.44$ and $j_2\approx 1.61$. For $ \Delta y=0.04$, the two first zeros are at  4.41  and 16.1. These are close to what we have at Figs.\,\ref{f5} and \ref{f6} although here we study an approximate solution for $\Delta x\ll 1$ whereas the figures  are results of ``exact" numerical integration.
  
Let us now consider the magnetic field $H$ applied parallel to the junction plane $(x,y)$. In general, the critical current should depend on both $H$ and $j$, $J_c=J_c(j,H)$. In the absence of injected currents,   $J_c(0,H)$ has a standard   shape with maximum at $H=0$. 
The presence of zeros of $J_c(j)$   for symmetric injection in zero field, has an important consequence. If the injected current $j_n$ is such that $J_c(j_n)=0$, application of the magnetic field will result in the pattern $ J_c(j_n,H)$ such that $ J_c(j_n,0)=0$, i.e., the curve  $ J_c(j_n,H)$ will have zero at $H=0$ instead of the standard maximum. This situation is similar to the famous case of ``0-$\pi$" junction, \cite{Lev}  however, here the injected currents cause a necessary phase shift. Precisely this situation has been seen in experiment \cite{Ust,Gold1} with symmetric injectors.  
 
 \subsection{Asymmetric injection}
\begin{figure}[h]
\begin{center}
\includegraphics[width=8.cm] {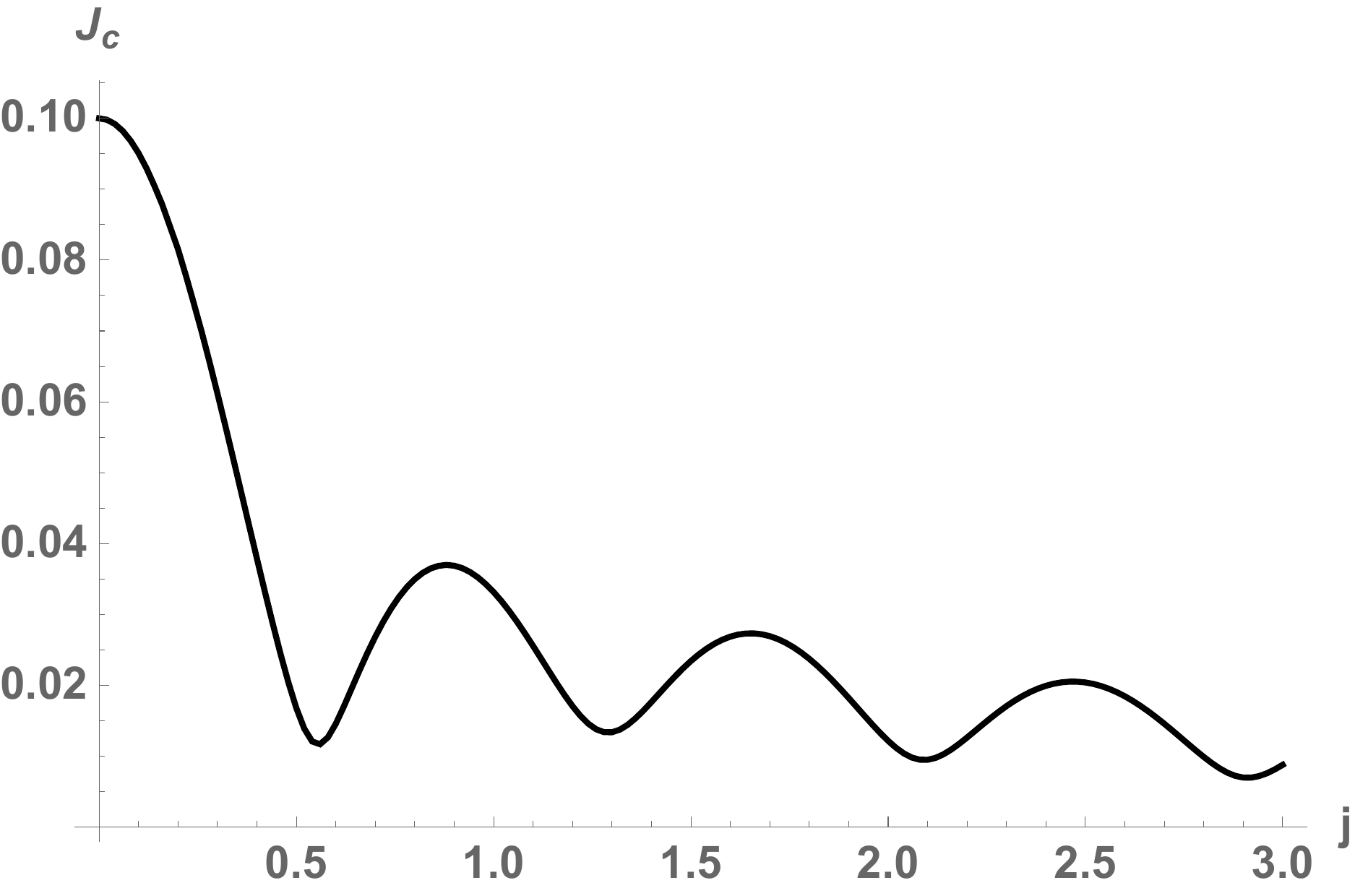} 
\caption{  $J_c(j)$ corresponding to the current distribution of Fig.\,\ref{f3} for the injection points $(0.2,0)$ and $(0,0.7)$.  
}
\label{f7}
\end{center}
\end{figure}

If the injection contacts are arranged asymmetrically as, e.g., at  Fig.\,\ref{f3},
 the minima of $J_c(j)$ do not reach zeros as shown in  Fig.\,\ref{f7}.
Without discussing a variety of asymmetric injections, we note that $J_c(j)$ of all of them have the property that their minima do not reach zeros. 

\section{Discussion}
In summary, we have shown that the current distribution in thin film samples small on the scale of the Pearl length $\Lambda=2\lambda^2/d$ can be found by solving the  Laplace equation for the stream function under boundary conditions  specified for injection sources at arbitrary points at  sample edges. When this film constitutes one of the Josephson junction banks, the contribution of the phase associated with injected currents  to the junction phase difference is proportional to the injected current $j=4\pi\Lambda I/c\phi_0$. Hence, the thinner the film (or the larger the Pearl $\Lambda$) the smaller  injected currents are needed for the same effect upon the junction  properties. The   critical Josephson current   $J_c(j)$, for certain (symmetric) infection geometries,  has zeros, the position of which scales as the inverse distance between the injection points. If $j_n$ is one of these zeros, application of a field $H$ parallel to the junction plane results in a pattern $J_c(H,j_n) $ with  zero at $H=0$ instead of a standard maximum, the property seen in experiments. \cite{Ust,Gold1}  \\

 The authors are grateful to E. Goldobin for sharing  experimental information and  many helpful discussions.  The Ames Laboratory is supported by the Department of Energy, Office of  Basic Energy Sciences, Division of Materials Sciences and Engineering under Contract No. DE-AC02-07CH11358.


\begin{thebibliography}{99}

 \bibitem{Lev} L. N. Bulaevskii, V. V. Kuzii and A. A. Sobyanin, Sol. St. Com, {\bf 25}, 1053  (1978).
 
 \bibitem{Kirtley}   C. C. Tsuei and J.R. Kirtley, Rev. Mod. Phys. {\bf 72}, 969 (2000).

 \bibitem{Mints-phi}R. G. Mints,  \prb {\bf 57}, R3221 (1998). 

\bibitem{Krasnov}T. Golod, A. Rydh, and V. M. Krasnov,  \prl {\bf 104},   227003 (2010).

  \bibitem{KM1} V. G. Kogan, R. G.  Mints,   \prb {\bf 89}, 014516 (2014).
 
  \bibitem{KM2} V. G. Kogan, R. G.  Mints,   Phys. C Superc. {\bf 502}, 58 (2014).
  
 \bibitem{Ust}A. V. Ustinov, Appl. Phys. Lett. {\bf 80}, 3153 (2002).


 \bibitem{Gold1}E. Goldobin, A. Sterck, T. Gaber, D. Koelle, and R. Kleiner, \prl {\bf 92}, 057005 (2004).


\bibitem{Gold3}A. Dewes, T. Gaber, D. Koelle, R. Kleiner, and E. Goldobin, \prl {\bf 101}, 247001 (2008).

 \bibitem{Morse}  P. M. Morse and H. Feshbach {\it Methods of Theoretical Physics}, McGraw-Hill, 1953. 

  \bibitem{privat}E. Goldobin, private communication.




 
%
\end{thebibliography}
\end{document}